\documentclass[12pt]{article}
\usepackage{graphicx}
\usepackage[cp1251]{inputenc}
\usepackage{rotating}
\begin{document}
 \noindent {\footnotesize\it Astronomy Reports, 2021, Vol. 65, No 6, pp. 498--506}
 \newcommand{\dif}{\textrm{d}}

 \noindent
 \begin{tabular}{llllllllllllllllllllllllllllllllllllllllllllll}
 & & & & & & & & & & & & & & & & & & & & & & & & & & & & & & & & & & & & & &\\\hline\hline
 \end{tabular}

 \vskip 0.5cm
  \centerline{\bf\Large A New Estimate of the Best Value for the Solar}
  \centerline{\bf\Large Galactocentric Distance}
 \bigskip
 \bigskip
  \centerline
 {
 V.V. Bobylev and A.T. Bajkova
 }
 \bigskip
 \centerline{\small \it
 Central (Pulkovo) Astronomical Observatory, Russian Academy of Sciences,}
 \centerline{\small \it Pulkovskoe shosse 65, St. Petersburg, 196140 Russia}
 \bigskip
 \bigskip
 \bigskip

 {
{\bf Abstract}---Using data from the literature, we made a list of individual estimates of the solar Galactocentric distance, which were performed after 2017 by different methods. These values have not yet been used to calculate the best value of mean $R_0$. For the sample containing 21 estimates, based on the standard approach, we found the weighted mean ${\overline R_0}=8.14$ kpc with the dispersion $\sigma=0.16$ kpc, and using the median statistics, we obtained the estimate $R_0=8.15\pm0.11$ kpc. For practical use, the value $R_0=8.1\pm0.1$ kpc can be recommended.
  }

\medskip DOI: 0.1134/S1063772921070015

 \section{INTRODUCTION}
The solar Galactocentric distance $R_0$ is one of the most important fundamental astronomical parameters, the exact knowledge of which is of great importance for astrophysics and cosmology. The ``standard'' values of this quantity recommended by the International Astronomical Union (IAU) differ markedly, amounting to $R_0=10$~kpc (IAU, 1964) and 8.5~kpc (IAU, 1986). Modern estimates give a value close to $8.0$~kpc [1--6].

There are various methods for assessing $R_0$, for which various types of classification have been proposed. Reid [7] divided all such measurements into three classes: direct, secondary, and indirect. Bland-Hotthorn and Gerhard [3] adhere to a close classification,
dividing all measurements into three classes: direct, model-dependent, and secondary. Nikiforov [1] proposed a special three-dimensional classification. He divided measurements into three classes based on the type of measurement, the type of the
$R_0$ estimate, and the type of reference objects.

The method for determining the absolute trigonometric parallax of an object located close to the Galactic center is truly straightforward. Based on VLBI observations of several maser sources in the Sgr B2 region, this method was used to estimate of $R_0=7.9^{+0.8}_{-0.7}$ 
kpc [8]. The dynamic parallax method is also highly accurate and reliable. From a joint analysis of the orbital motion of 28 stars around the central supermassive black hole, Gillessen et al. [9] found $R_0=8.33\pm0.31$ kpc by this method with a simultaneous
estimate of the black hole mass $(4.31\pm0.38)\times10^6~M_\odot$. During high-precision
astrometric observations of these stars, only one star, S2, with an orbital period of about 16 years, has completed a full revolution. The orbital periods of the remaining 27 stars are 45 years or more, up to 1000 years [9]. Therefore, in order to refine the estimate,
it is the motion of the star S2 that is usually analyzed [10--13]. To date, the application of this method makes it possible to estimate with a relative error of about 0.3\% [12, 13].

Variable stars---classical Cepheids, type II Cepheids and RR Lyr variables---are important for the $R_0$ estimate. High accuracy of distance estimates to Cepheids is possible due to the period--luminosity relationships [14, 15] and the period--Vesenheit function
[16, 17]. These relationships are well calibrated using high-precision trigonometric stellar parallaxes [18]. Their use makes it possible to estimate distances to Cepheids with relative errors less than 10\% [19, 20]. Moreover, according to [21], errors in the distances to
Cepheids are $\sim$5\%. Although a local systematization is
not excluded, where strong differences in the metallicity
of stars, a non-standard ratio of total to selective
absorption, etc. According to estimates by some
authors [22], the distances to RR Lyr variables can
currently be measured on average with relative errors
of about 4\%. For the $R_0$ estimate, it is required to identify
a group of such variable stars that are located in the Galactic disk, bulge, or halo, and are symmetrically distributed relative to the center of the Galaxy.

Note that the maser sources have measured trigonometric parallaxes [23, 24]. At present, the results of radio observations using the VLBI method for more
than 200 such sources have been published. The distances
to them were measured on average with relative
errors of about 6\%. The estimates based on these data
were obtained mainly by the kinematic method [23--25]. Nikiforov and Veselova [26] proposed an interesting method for estimating $R_0$ from the distribution of masers in spiral arms.

The aim of this paper is to obtain a new estimate of the mean $R_0$ from the analysis of the latest individual determinations. For practical use, it is important to know the most probable value of the error in determining $R_0$ that we plan to estimate. The objective
value of such an error is necessary, in particular, when evaluating the linear Galactic rotation velocity $V_0$ from the measured value of the angular velocity of its rotation $\Omega_0$ ($V_0=R_0\Omega_0$), as well as when evaluating, for example, the uncertainty of the Oort constants $A=-0.5\Omega'_0 R_0$ and $B=-\Omega_0+A$.

\section{DATA}
To date, there are a number of studies in which the mean value of the ``best'' distance $R_0$ is derived on the basis of individual determinations of this value
obtained by independent methods over a certain long
time interval. Such results are shown in Table 1. The
first column of the table gives the mean $R_0$ with the
error estimate, the second column shows the number
of independent measurements used to calculate the mean, the third column indicates the time interval during which individual estimates $R_0$ were obtained, and the last column provides a reference to the authors of the individual estimate.

Table 1 shows the mean values of $R_0$ obtained from the analysis of individual estimates published over a time interval from 6 to 20 years. Particularly noteworthy
is the result from [5], in which 162 individual estimates
were analyzed over a 100-year time interval. It
can be seen that all the mean values of $R_0$ presented in
this table are in very good agreement with each other. Note that not all results are completely independent. For example, the sets of initial data in [4] and [6] are practically common. But these authors use different statistical methods to analyze the data.

The main conclusion that can be drawn from the analysis of Table 1 is that the value of $R_0$ is close to 8.0 kpc. And this value is very different from
$R_0=8.5$~kpc recommended by the IAU in 1986.

The first column of the table gives a value of $R_0$ with an error estimate corresponding to the level $1\sigma,$ the second column contains the type of stars used or the method of $R_0$ estimate, the third column contains the number of objects used for the calculation of $R_0$, and the last column provides a reference to the authors of the $R_0$ estimate.

Note that some authors estimate both the statistical and systematic error of the result. For example, in [5], $R_0=8.3$~kpc was obtained with an error $\pm0.2\,(stat.)$ and $\pm0.4\,(syst.).$ Further, we will write it as $R_0=8.3\pm0.2\pm0.4$~kpc. In the present paper, when refer to an estimate, we usually take the systematic error as an error. In this case, as for $R_0=7.93\pm0.13\pm0.04$~kpc [11], we take the maximum error value.

Table 2 contains 21 estimates of $R_0$. Four estimates were obtained from the analysis of the orbital motion of the star S2 around the supermassive black hole in
the center of the Galaxy [10, 11, 13, 27]. In all these
four cases, there are partially common astrometric
measurements of the positions of the star S2, but there
are differences, for example, in the number and quality
of the radial velocities of this star. Indeed, in the
studies of the GRAVITY collaboration [12, 27], observations
were carried out at the European Southern
Observatory in Chile using an optical interferometer
Very Large Telescope Interferometer (VLTI). The
results of another team [11, 13] were obtained mainly on the basis of observations at the Keck Observatory located on the Mauna Kea Mountain in Hawaii.

Quite recently, paper [12] was published, where the estimate of $R_0=8.178\pm0.013\pm0.022$~kpc was obtained. But the most recent publication of this
group [27] shows the presence of instrumental aberrations. Therefore, all previous estimates of the collaboration, starting from 2018 (in particular, the result of [12]), were revised, and a value of $R_0=8.275\pm0.009\pm0.033$~kpc was proposed. This is what we use
in this paper.

As observed from all these results, the estimates [27, 13] were obtained with the smallest random errors. Therefore, they should have the greatest weights when calculating the weighted mean. Note that the random errors of the presented $R_0$ estimates
differ by an order of magnitude. In this case, a weighing system must be used.

\begin{table}[t]
\caption{Results of determining the ``best value'' of mean $R_0$}
\begin{center}
\begin{tabular}{|l|c|c|c|r|}
\hline
 $R_0\pm\varepsilon_R (1\sigma),$ kpc   & $n$ & Years & Reference \\\hline
 $8.0\pm0.5 $ &  37 & 1972-1993 & [7] \\
 $7.9\pm0.17$ &  65 & 1974-2003 & [1] \\
 $8.0\pm0.25$ &  53 & 1992-2011 & [2] \\
 $8.2\pm0.1 $ &  26 & 2009-2014 & [3] \\
 $8.0\pm0.2 $ &  27 & 2012-2017 & [4] \\
 $8.3\pm0.4 $ & 162 & 1927-2017 & [5] \\
 $8.0\pm0.15$ &  28 & 2011-2017 & [6] \\
\hline
\end{tabular}
\label{t-1}
\end{center}
\end{table}
%%%%%%%%%%%%%%%%
\begin{table}[t]
\caption{Individual results of determining the distance $R_0$}
\begin{center}
\begin{tabular}{|c|c|c|c|r|}\hline
      $R_0\pm\varepsilon_R (1\sigma)$ & Star type/Method & $n$ & Reference \\
                                 kpc  &                 &     &        \\\hline
 $7.99 \pm0.49 $ &   RR Lyr type variables &        850 & [31] \\ %Muhie,Dam
 $8.27 \pm0.10 $ &      classical Cepheids &  $\sim$800 & [43] \\ %Bob,Rast
 $8.275\pm0.033$ &                      S2 &          1 & [27] \\ %Abuter
 $8.15 \pm0.12 $ &  masers and radio stars &        256 & [25] \\ %Bob,Kris
 $8.2  \pm0.6  $ &       blue giant branch & $\sim$2500 & [36] \\ %Utkin
 $8.28 \pm0.14 $ &   RR Lyr type variables &      16221 & [30] \\ %Griv
 $7.92 \pm0.30 $ &  masers of VERA program &    99 & [24] \\ % Hirota 2020
 $8.15 \pm0.15 $ & masers of BeSSeL program &   199 & [23] \\ % Reid 2019
 $7.971\pm0.032$ &                      S2 &     1 & [13] \\ %Do,S0-2
 $8.1  \pm0.2  $ &   RR Lyr type variables &  2016 & [29] \\ %
 $8.2  \pm0.1  $ &      classical Cepheids &   218 & [42] \\ %
 $7.9  \pm 0.3 $ &          Mira variables &  1863 & [34] \\ %Qin,
 $7.6  \pm0.7  $ &       globular clusters &   119 & [45] \\ %
 $8.05\pm0.024~* $ & RR Lyr type variables & 960 & [28] \\ %
 $8.30 \pm0.36 $ &   RR Lyr type variables &  4194 & [22] \\ %MajR0=8.30±0.36
 $8.46 \pm0.11 $ &              Cepheid II &   894 & [33] \\ %
 $8.10 \pm0.22 $ &      classical Cepheids &     4 & [35] \\ % Chen
 $7.93 \pm0.13 $ &                      S2 &     1 & [11] \\ % Chu
 $8.8  \pm 0.5 $ & sections of spiral arms &  2 & [26] \\ %Nikif
 $8.34 \pm 0.41$ &              Cepheid II &   264 & [32] \\ %Bhardwaj
 $8.32 \pm 0.14$ &                      S2 &     1 & [10] \\ %Gillessen
   \hline
\end{tabular}
\label{t-2}
\end{center}
{\def\baselinestretch{1}\normalsize\small
 (*)~---the accuracy of the estimate is strongly overestimated here; in what follows, for this result, we will use the value of the random error of 0.11~kpc.
 }
\end{table}
%%%%%%%%%%%%%%%%%%%%%%%%

Nine $R_0$ estimates are derived from an analysis of
the spatial distribution of variable stars. Four estimates
were obtained for RR Lyr type variables [22, 28--31],
two for type II Cepheids [32, 33], one for Mira variables
[34], and one for classical Cepheids [35]. In
terms of the similarity of the approach, they are similar
to the result obtained for the blue-giant branch stars
[36].

Type II Cepheids were used in [32, 33]. They are
low-mass stars, poor in metals. They are found in
globular clusters, the Galactic disk and the Galactic
bulge. Type II Cepheids are more than magnitude
fainter than classical Cepheids with similar periods,
and follow a slightly different period-luminosity relationship.
In [32], data from the Optical Gravitational
Lensing Experiment (OGLE-III [40]) and VISTA
Variables in the Via Lactea (VVV [41]) reviews are
combined. We selected 264 stars with good quality
light curves. In [33], the same team of authors
repeated the study using a much larger sample of
Cepheids. The estimates were obtained on the
assumption that the selected stars are distributed symmetrically
about the Galactic center.

RR Lyr variables belong to the horizontal giant branch in the Hertzsprung–Russell diagram. These are old stars belonging to Population II, containing few heavy elements and located in the spherical subsystem of the Galaxy. Commonly they found in globular clusters. In [22], the selection of RR Lyr variables was carried out from the VVV survey [41]. Individual
distances to these stars were estimated using near infrared photometry ($J,H,K_s$). $R_0$ was calculated using a high-latitude subsample of selected stars $(|b|>4^\circ).$ In
[28], candidates were also selected from the VVV catalog [41], but the individual distances to stars were estimated by them using other calibrations. In [29],
the SEKBO (Southern Edgeworth–Kuiper Belt Object [42]) survey was used to select RR Lyr variables. In [30], reviews of the OGLE program [43] were used for these purposes. In [31], new spectral and photometric observations with the Southern African Large Telescope (SALT) and a new calibration for 850 RR Lyr stars were used for the $R_0$ estimate.
According to estimates [22], the dispersion in determining the absolute magnitude of RR Lyr stars is $M_{K_s}=0.08^m$. This means that, on average, the random error in determining the individual distance to such variable stars is about 4\%.

In [35], a sample of 55 classical Cepheids belonging to the Galactic bulge was studied. Considering absorption, we used their photometric observations in seven
ranges $J,$ $H,$ $K_s,$ 3.6 $\mu$m, 4.5 $\mu$m, 5.8 $\mu$m, and 8.0 $\mu$m.
In these ranges, the interstellar extinction value is significantly
lower than in the optical range. As is known,
the estimate of the individual distance to the star, the
Cepheid in particular, strongly depends on the correct
accounting for the absorption. As a result, according
to [35], the error in estimating the distances to these
Cepheids averaged about 4--5\%. They are unevenly
distributed relative to the center of the Galaxy. The
basic mass is located outside the center of the Galaxy
at a mean distance of about 12 kpc. And only 4 stars
are located in the center of the Galaxy, according to
which the $R_0$ estimate was obtained.

A large sample of Mira variables was studied in [34]. These stars are pulsating variables that are in the late evolutionary stages of the asymptotic giant
branch. They are characterized by long periods of pulsation
(over 100 days) and high near infrared and bolometric
luminosities. They belong to the Galactic halo.
In [34], a large sample was formed, compiled from
data from several observational programs (SAAO, MACHO, and OGLE). Calibration and absorption accounting was performed using photometry $J,H,K_s$ from the 2MASS catalog [44]. The $R_0$ estimate was obtained on the assumption that the selected stars are
distributed symmetrically about the Galactic center.

In [36], stars of the blue giant branch, which are also halo objects, were used. The $R_0$ estimate was obtained by analyzing their kinematics based on the
statistical parallax method.

Compared to the random errors of $R_0$ obtained
from other samples of RR Lyr stars, the random errors
in [28] are too small. Therefore, when calculating the
weighted mean, we increased them by a factor of 10,
i.e., made them comparable with the estimates
[22, 29].

Five estimates of $R_0$ were obtained from an analysis
of the kinematics of maser sources with measured trigonometric
parallaxes [23, 24], maser sources and radio
stars [25], as well as from various samples of classical
Cepheids [37, 38]. Data on 199 masers observed at
different frequencies (methanol at 8.4 GHz and
H2O-masers at 22 GHz) in the framework of the Bar
and Spiral Structure Legacy Survey (BeSSeL\footnote{http://bessel.vlbi-astrometry.org}) project are described in [23]. In [24], 99 maser sources were
analyzed, which were observed at a frequency of
22 GHz in the framework of the Japanese VLBI
Exploration of Radio Astrometry (VERA\footnote{http://veraserver.mtk.nao.ac.jp}) program.
Note that the higher the frequency, the more accurately
the VLBI observations of parallaxes and proper
motions of radio objects are obtained. Most of the data
from [24] was included in the sample [23]. In [25], the
sample [23] was supplemented by VLBI observations
of radio stars, which, however, lie very close to the Sun
in the Gould Belt region. In [38], the latest data on
classical Cepheids from [45, 21] were used. These
Cepheids belong to the disk component of the Galaxy
and are distributed practically over the entire disk. In
all five cases noted here, $R_0$ entered as an unknown
when solving the basic kinematic equations describing
the Galactic rotation. According to Reid’s classification
[7], this approach belongs to indirect methods,
and according to the classification of Bland-Hotthorn
and Gerhard [3], it refers to model-dependent methods.

The $R_0$ estimate was also obtained by a kinematic
method in [39]. For this, the spatial velocities of 119
globular clusters were analyzed. Moreover, for them,
the original values of proper motions were calculated,
obtained with an epoch difference of about 65 years.

Finally, Table 2 presents the $R_0$ estimate obtained
in [26] from an analysis of the distribution of masers
with measured trigonometric parallaxes in the spiral
arms. They used masers located in the two sections of
the spiral arms closest to the Sun--Perseus and
Carina--Sagittarius. So far, however, there is little data
for a confident application of the method, so the
$R_0$ estimation error turned out to be large.

For statistical analysis and comparison of the
results obtained with other authors, we formed a sample
of 35 measurements performed during 2011--2017.
These data are described in [6], which also contains
some results from [4], in which the authors of [6] did
not include in their sample. Thus, we have created a
more complete list of measurements for this time
period.

At the same time, the very inaccurate estimate $R_0=7.6\pm1.35$~kpc obtained by the statistical method for planetary nebulae in [46] was not included in the sample. Finally, one of our results obtained for 73 masers, $R_0=8.3\pm0.3$~kpc [47], was added to the
sample.

All individual $R_0$ estimates used in this paper, depending on the year of publication, are shown in Fig. 1. They cover a ten-year range from 2011 to 2021,
where the so-called ``bandwagon'' effect is completely
invisible. This effect was noticed by Reid [7]. It manifests
itself as a tendency to obtain a new estimate close
to the current standard value.

\begin{table}[t]
\caption{Estimates of the mean $R_0$ and its errors, obtained by the standard method}
\begin{center}
\begin{tabular}{|c|c|c|c|c|c|}\hline
 $n$ & ${\overline R_0}$ & Dispersion & Error of mean & Weight $(w)$ & Marks in   \\
     &               kpc &       kpc  &   kpc &       & Fig.~\ref{fig-1} \\\hline
 21 & 8.157 & 0.239 & 0.052 &  1           & Red \\
 21 & 8.139 & 0.157 & 0.034 & $1/\varepsilon_R^2$ & Red \\\hline

 35 & 7.977 & 0.422 & 0.071 &  1           & Blue \\
 35 & 7.973 & 0.304 & 0.051 & $1/\varepsilon_R^2$ & Blue \\\hline

 56 & 8.044 & 0.374 & 0.050 &  1           & All \\
 56 & 8.090 & 0.225 & 0.030 & $1/\varepsilon_R^2$ & All \\\hline
\end{tabular}
\label{t-rezults}
\end{center}
\end{table}
%%%%%%%%%%%%%%%%%%%%%%%%
%%%%%%%%%%%%%%%%%%%%%%%% FIG.1:
\begin{figure}[t]
{ \begin{center}
    \includegraphics[width=0.95\textwidth]{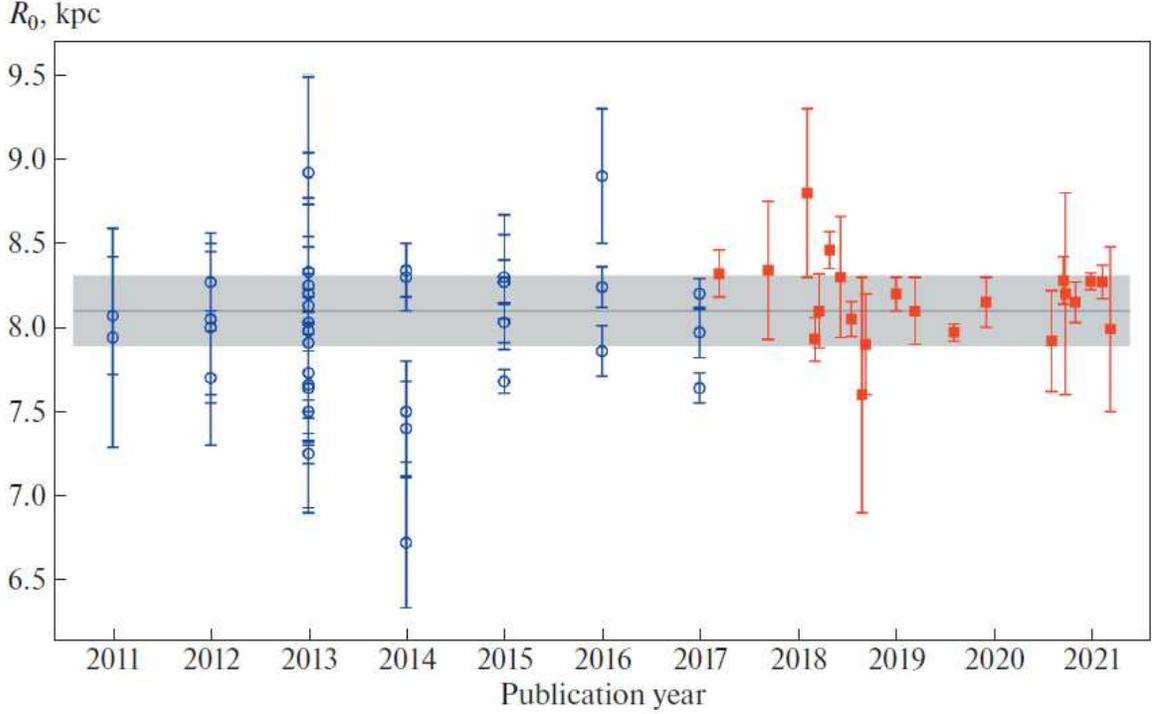}
\caption{The values of $R_0$ obtained by various authors in the last decade: data from the review [6] are indicated by open blue circles and data from Table 2 of this paper are presented by red squares. The mean ${\overline R_0}=8.090$~kpc (gray horizontal line) and the confidence region corresponding to the dispersion $1\sigma=0.225$~kpc (gray fill) are shown.}
  \label{fig-1}
  \end{center}
  }
\end{figure}
%%%%%%%%%%%%%%%%
%%%%%%%%%%%%%%%%%%%%%%%% FIG.2:
\begin{figure}[t]
{ \begin{center}
   \includegraphics[width=0.65\textwidth]{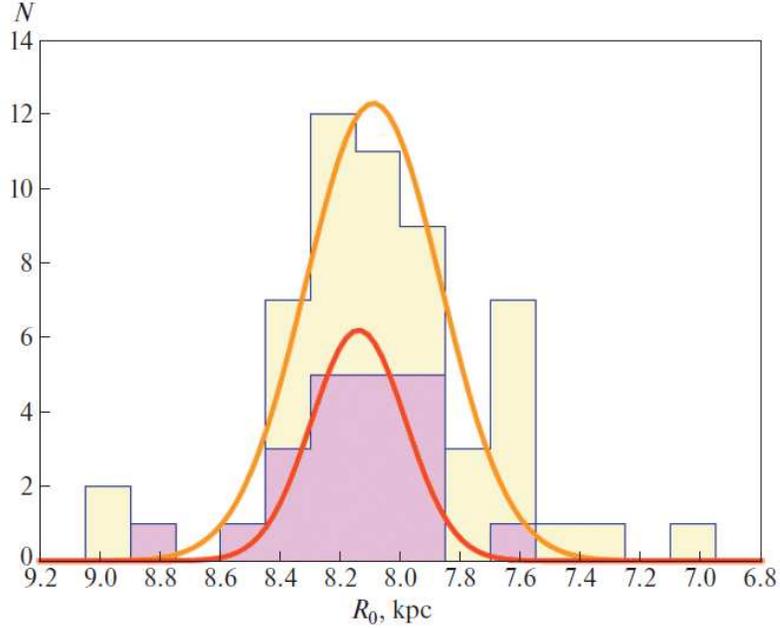}
\caption{Histograms of distribution for samples consisting of 56 values (light fill) and 21 values of $R_0$ (dark fill) with corresponding Gaussians, see the text for details.}
  \label{fig-2}
  \end{center}
  }
\end{figure}
%%%%%%%%%%%%%%%%
%%%%%%%%%%%%%%%%%%%%%%%% FIG.3:
\begin{figure}[t]
{ \begin{center}
  \includegraphics[width=0.95\textwidth]{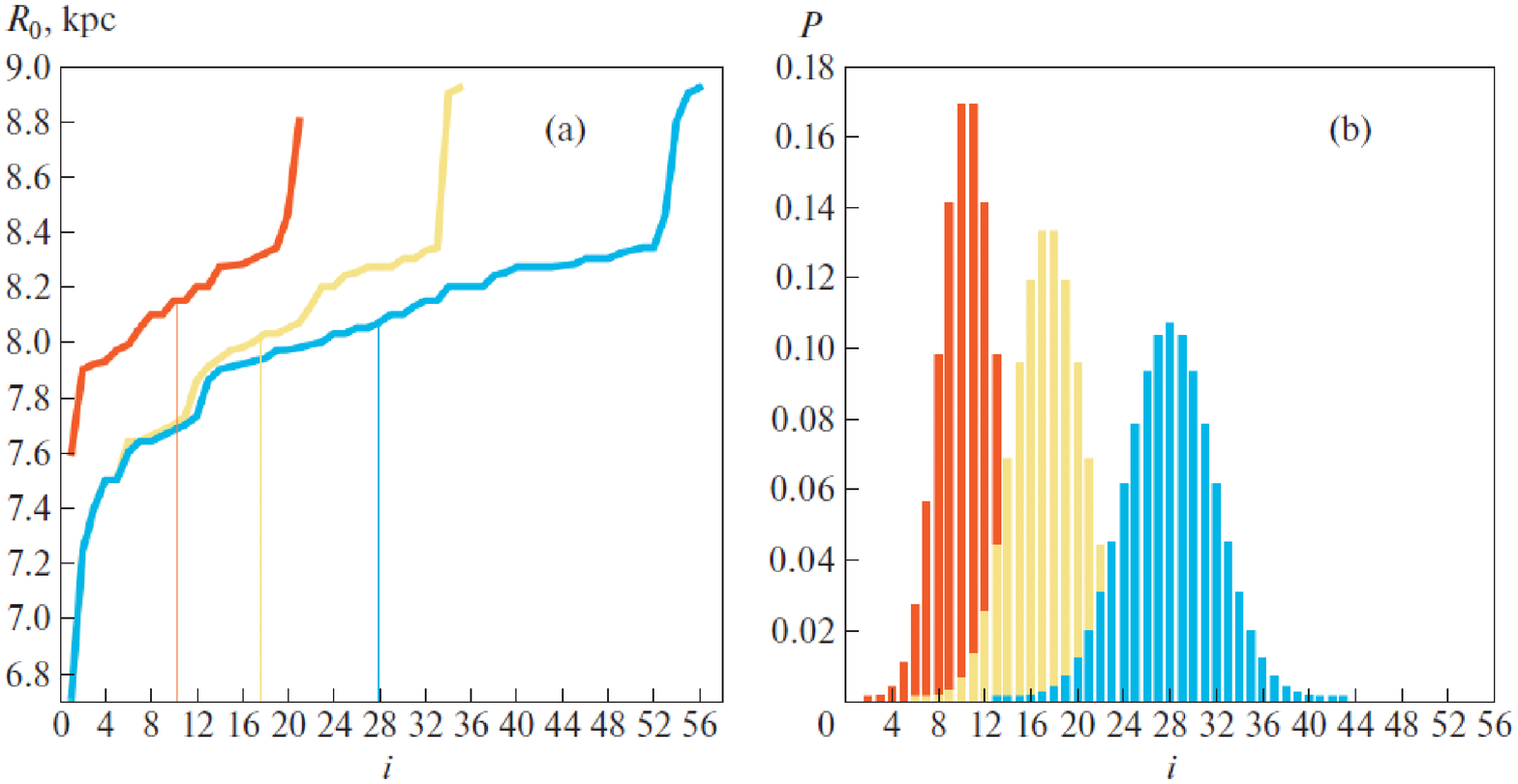}
\caption{Ordered sequences of measurements showing median values (a) and probability distributions (b) versus sequence number $i$ for three samples.
  }
  \label{fig-3}
  \end{center}
  }
\end{figure}
%%%%%%%%%%%%%%%%

\section{RESULTS AND CONCLUSIONS}
 \subsection{Traditional Approach}
This approach consists in calculating the arithmetic
mean, the weighted mean of the parameter $R_0$, as
well as the error estimates, based on the assumption of
their Gaussian distribution. (Note that the arithmetic
mean is obtained with unit weights.)

Table 3 shows the values of the mean distance $R_0.$
The calculations were carried out for three samples:
first, for a sample of 21 individual assessments, which
are given in Table 2; second, for a sample of 35 individual
estimates, which we formed according to data
from [6]; and third, for a pooled sample of 56 individual
estimates, which contains the results from 2011 to
2021. The mean value of ${\overline R_0}$ is calculated in accordance
with the well-known formula:
 \begin{equation}
 \renewcommand{\arraystretch}{1.8}
 %\begin{array}{lll}
 {\overline R_0}={{\sum^n_{i=1} w_i{R_0}_i}\over{\sum^n_{i=1} w_i}},
 \label{Average}
 %\end{array}
 \end{equation}
where $n$ is the number of measurements, $w_i$ is the weight of the $i$-th dimension, as can be seen from the Table 3, weights of the following two types are used: $w_i=1$ and $w_i=1/\varepsilon_R^2$. The dispersion of the estimate is calculated according to the formula
 \begin{equation}
 \sigma =
  \sqrt{{\sum^n_{i=1} w_i({R_0}_i-\overline R_0)^2}\over
        {\sum^n_{i=1} w_i}}
 \label{er-1}
 \end{equation}
Then, the error of the mean (or weighted mean)
 \begin{equation}
  \varepsilon = {\sigma\over\sqrt n}.
 \label{er-2}
 \end{equation}
As we can see from Table 3, for any data segmentation into subsamples, we obtain the mean value of ${\overline R_0}$ in a very narrow range of 8.0--8.2 kpc. The dispersion values
are very different. The type of weighting coefficients noticeably affects both the dispersion and the error of the mean.

In our case, we have a slightly larger number of members in the sample $n=35$ from the 2011--2017 interval compared with the sample analyzed by Camarillo et al. [6], where $n=28$. Despite this, the result we found, ${\overline R_0}=7.973\pm0.051$~kpc (fourth row from the top in Table 3), is in good agreement with the estimate [6] 
 ${\overline R_0}=7.93\pm0.03$~kpc (weighted mean, with weights of the type $w_i=1/\varepsilon_R^2$ and dispersion $\sigma=0.40$~kpc). It
can also be noted that our estimate is in good agreement with the result of [2], which was obtained on the basis of a similar ${\overline R_0}=7.967\pm0.048$~kpc (weighted
mean error) method from the analysis of a completely
different set of individual estimates of $R_0$.

We consider the weighted mean ${\overline R_0}=8.139\pm0.034$~kpc found for the sample from the 2017--2021 interval to be the most interesting ($n=21$, second line from
the top in Table 3). This estimate is new, based on
independent data. It can be seen that the variance values
in this sample favorably differ in the smaller direction
from the dispersion values calculated for the previous
time period.

Figure 1 shows the mean ${\overline R_0}=8.090$~kpc value found from a sample of 56 measurements and indicates the confidence region corresponding to the standard deviation ($1\sigma=0.225$~kpc) found for this entire sample. At the same time, it is clearly seen that the confidence region for the red points should be approximately twice as narrow.

Figure 2 shows two histograms. One histogram is based on a sample of 56 individual determinations of $R_0$ (light shading) with mean value ${\overline R_0}=8.090$~kpc and $\sigma=0.225$~kpc. The other is based on a sample of 21 individual definitions of $R_0$ (dark shading) with mean value ${\overline R_0}=8.139$~kpc and $\sigma=0.157$~kpc.
Camarillo et al. [6] noted that the distribution of the 28 estimates of $R_0$ used in [6] is wider than the Gaussian distribution and has other small deviations from
the Gaussian. Such manifestations can be seen in Fig. 2 (the distribution wings are broadened, the secondary maximum at $R\sim7.6$~kpc in the distribution of 56 estimates), although they are not large.

In [6], the non-Gaussianity in the distribution of errors was proved using the Student distribution. Since the approach used to determine the most probable value of the error of the mean $R_0$ in [6] is of interest, we also decided to apply it.

 \subsection{Median Statistics}
A description of the approach can be found in [6, 48, 49], in which it was applied when searching for the most probable values of some astronomical parameters. Authors of these papers call their approach ``median statistics''. Moreover, it is completely different from the median method used, for example, in
papers [2, 50].

The median is the center value in a sorted dataset
that divides the dataset into two halves, each with an
equal number of elements. Median statistics assume
statistical independence of all measurements and the
absence of systematic errors. It does not use measurement
errors, which is an advantage if the errors are
non-Gaussian or estimated incorrectly. The payback
for it is that the central median estimate has relatively
large uncertainty. To find errors related to the median
value, we follow [48].

For a dataset consisting of $N$ independent measurements $M_i$, we determine the probability $P$ of the median $M_{med}$, lying between $M_i$ and $M_{i+1}$ as the binomial
distribution:
 \begin{equation}
  P= {{2^{-N} N!}\over {i!(N-i)!}}.
 \label{er-2-1}
 \end{equation}
Median errors $M_{med}$ are defined as follows. From the value of $M_{med}$, which has the highest probability, integration is performed in both directions. The stop occurs when the cumulative probability reaches 0.6827 of the total probability, which corresponds to the standard deviation $1\sigma$. Next, the difference between the median $M_{med}$ and the two values $M$, corresponding to the ends of the integrals, is taken to obtain an error of one standard deviation, $1\sigma^+_-$. We then continue this integration until the cumulative probability reaches 0.9545 of the total probability to get the standard deviation errors in $2\sigma^+_-$. Note that the distribution does not have to be symmetric, so the bounding values $2\sigma^+_-$ are not necessarily twice the value $1\sigma^+_-$.

Table 4 shows the $R_0$ estimates and its errors obtained by the method just described. As shown in [6], the error estimates are not symmetric. Therefore,
Table 4 gives estimates of $R_0$ for both the level $1\sigma^+_-$ and
$2\sigma^+_-$. Note that the mean value is not calculated in this
method, but the median value is taken from the available
ordered list (therefore, we do not put the average
sign here).

\begin{table}[t]
\caption{Estimates of $R_0$ and its errors obtained by the
``median statistics'' method}
\begin{center}
\begin{tabular}{|c|c|c|c|c|c|}\hline
 $n$ & $R_0\pm1\sigma$ & $R_0\pm2\sigma$ & Interval $\pm1\sigma$ & Interval $\pm2\sigma$\\
     &             kpc &      kpc  & kpc & kpc \\\hline
 21 & $8.15^{+0.12}_{-0.10}$ & $8.15^{+0.13}_{-0.18}$ & 8.05--8.27 & 7.97--8.28 \\

 35 & $8.03^{+0.10}_{-0.12}$ & $8.03^{+0.17}_{-0.33}$ & 7.91--8.13 & 7.70--8.20 \\

 56 & $8.08^{+0.12}_{-0.09}$ & $8.08^{+0.12}_{-0.11}$ & 7.99--8.20 & 7.97--8.20 \\\hline
\end{tabular}
\label{t-rez-Median}
\end{center}
\end{table}
%%%%%%%%%%%%%%%%%%%%%%%%

The second line in the table is given for comparison
with the result of [6], in which a sample of 28 estimates
using the median statistics, $R_0=7.96^{+0.11}_{-0.23}$, was found
for the $1\sigma$ level and $R_0=7.96^{+0.24}_{-0.30}$ for the $2\sigma$ level.
Camarillo et al. [6] estimates were made from very
close samples.

The method is illustrated in Fig. 3. Fig. 3a shows
three ascending measurement sequences for three
samples, which contain 21 (red), 35 (sandy), and
56 (blue) measurements, respectively. The vertical
lines in this figure represent the median value. The
probability distribution $P$ (see expression (4)) for the
same samples is shown in Fig.~3b.

The most interesting result in Table 4, of course, is
the result obtained from the sample containing 21 estimates.
Assuming that the errors are symmetric, we can
write $R_0=8.15\pm0.11$~kpc. As we can see from Fig. 3a,
the sequence for a sample of 56 measurements (blue
line) has the smallest slope to the horizontal axis compared
to the other two sequences. Therefore, for this
sample, close errors were obtained, both for the level
$1\sigma^+_-$and $2\sigma^+_-$.

According to the second and last lines of the Table 3,
we have the most probable mean value $R_0=8.1$~kpc.
As the most probable estimate of the error of $R_0$, one
can take $0.1$~kpc value, which is in agreement with
both the value obtained using the standard approach
and using the median statistics. Then, for practical
use, the value $R_0=8.1\pm0.1$~kpc can be recommended.
Note that this value is in good agreement
with the results given in Table 1.

\section{CONCLUSIONS}
A statistical analysis of the estimates of the Galactocentric distance $R_0$ is performed. For this, we used the results obtained by various authors over the past
decade, from 2011 to 2021. For this entire sample containing 56 measurements, based on the standard approach, we found the weighted mean ${\overline R_0}=8.090$~kpc
with the dispersion $\sigma=0.225$~kpc and the error of the
weighted mean $\varepsilon_R =0.030$~kpc. For the same sample,
on the basis of median statistics, we found $R_0=8.08\pm0.10$~kpc.

Our list contains 21 individual estimates of $R_0$ since 2017. These results have not yet been used by anyone to calculate the best value of mean $R_0$. For this sample,
on the basis of the standard approach, we found a
weighted mean ${\overline R_0}=8.14$~kpc with a dispersion
$\sigma=0.16$~kpc and an error of the weighted mean $\varepsilon_R =0.03$~kpc. For the same sample, containing 21 estimates, based on the median statistics, we found $R_0=8.15\pm0.11$~kpc under the assumption of symmetry of errors.

As we have already noted, the median statistics
does not give a completely accurate mean value (the
median is assigned from the available list of measurements),
but it estimates the errors of the result well.
Therefore, we use the combined result. Namely, we
take the weighted mean $R_0$ and the errors are estimated
based on the median statistics.

As a result, we came to the conclusion that it is possible to recommend the value $R_0=8.1\pm0.1$~kpc for practical usage.

 \medskip\subsubsection*{REFERENCES}

 {\small
 \quad 
 ~1. I. I. Nikiforov, in Order and Chaos in Stellar and Planetary
Systems, Proceedings of the Conference, August 17--24, 2003, St. Petersburg, Russia, Ed. by G. G. Byrd, K. V. Kholshevnikov, A. A. Myllari, I. I. Nikiforov, and
V. V. Orlov, ASP Conf. Ser. 316, 199 (2004).

2. Z. Malkin, in Advancing the Physics of Cosmic Distances,
Ed. by R. de Grijs and G. Bono, Proc. IAU Symp. 289,
406 (2013).

3. J. Bland-Hawthorn and O. Gerhard, Ann. Rev. Astron. Astrophys. 54, 529 (2016).

4. J. P. Vall\'ee, Astrophys. Space Sci. 362, 79 (2017).

5. R. de Grijs and G. Bono, Astrophys. J. Suppl. 232, 22
(2017).

6. T. Camarillo, V. Mathur, T. Mitchell, and B. Ratra,
Publ. Astron. Soc. Pacif. 130, 4101 (2018).

7. M. J. Reid, Ann. Rev. Astron. Astrophys. 31, 345
(1993).

8. M. J. Reid, K. M. Menten, X. W. Zheng, A. Brunthaler,
and Y. Xu, Astrophys. J. 705, 1548 (2009).

9. S. Gillessen, F. Eisenhauer, S. Trippe, T. Alexander,
R. Genzel, F. Martins, and T. Ott, Astrophys. J. 692,
1075 (2009).

10. S. Gillessen, P. M. Plewa, F. Eisenhauer, R. Sari, et al.,
Astrophys. J. 837, 30 (2017).

11. D. S. Chu, T. Do, A. Hees, A. Ghez, et al., Astrophys.
J. 854, 12 (2018).

12. R. Abuter, A. Amorim, M. Baub\"ock, J. P. Berger, et al.,
Astron. Astrophys. 625, L10 (2019).

13. T. Do, A. Hees, A. Ghez, G. D. Martinez, et al., Science
(Washington, DC, U. S.) 365, 664 (2019).

14. H. S. Leavitt, Ann. Harvard College Observ. 60, 87 (1908).

15. H. S. Leavitt and E. C. Pickering, Harvard College Observ.
Circ. 173, 1 (1912).

16. B. F. Madore, Astrophys. J. 253, 575 (1982).

17. F. Caputo, M. Marconi, and I. Musella, Astron. Astrophys.
354, 610 (2000).

18. V. Ripepi, R. Molinaro, I. Musella, M. Marconi, S. Leccia, and L. Eyer, Astron. Astrophys. 625, A14 (2019).

19. L. N. Berdnikov, A. K. Dambis, and O. V. Vozyakova,
Astron. Astrophys. Suppl. Ser. 143, 211 (2000).

20. A. Sandage and G. A. Tammann, Ann. Rev. Astron.
Astrophys. 44, 93 (2006).

21. D. M. Skowron, J. Skowron, P. Mr\'oz, A. Udalski, et al., Science (Washington, DC, U. S.) 365, 478 (2019).

22. D. Majaess, I. D\'ek\'any, G. Hajdu, D. Minniti, D. Turner,
and W. Gieren, Astrophys. Space Sci. 363, 127 (2018).

23. M. J. Reid, K. M. Menten, A. Brunthaler, X. W. Zheng,
et al., Astrophys. J. 885, 131 (2019).

24. T. Hirota, T. Nagayama, M. Honma, Y. Adachi, et al.,
Publ. Astron. Soc. Jpn. 72, 50 (2020).

25. V. V. Bobylev, O. I. Krisanova, and A. T. Bajkova,
Astron. Lett. 46, 439 (2020).

26. I. I. Nikiforov and A. V. Veselova, Astron. Lett. 44, 81 (2018).

27. R. Abuter, A. Amorim, M. Baub\"ock, J. P. Berger, et al.,
arXiv: 2101.12098 [astro-ph.GA] (2021).

28. R. Contreras Ramos, D. Minniti, F. Gran, M. Zoccali,
et al., Astrophys. J. 863, 79 (2018).

29. E. Griv, M. Gedalin, and I.-G. Jiang, Mon. Not. R.
Astron. Soc. 484, 218 (2019).

30. E. Griv, M. Gedalin, P. Pietrukowicz, D. Majaess, and
I.-G. Jiang, Mon. Not. R. Astron. Soc. 499, 1091
(2020).

31. T. D. Muhie, A. K. Dambis, L. N. Berdnikov, A. Y. Kniazev,
and E. K. Grebel, Mon. Not. R. Astron. Soc. 502,
4074 (2021); arXiv: 2101.03899 [astro-ph.GA].

32. A. Bhardwaj, M. Rejkuba, D. Minniti, F. Surot, et al.,
Astron. Astrophys. 605, A100 (2017).

33. V. F. Braga, A. Bhardwaj, R. Contreras Ramos, D. Minniti,
G. Bono, R. de Grijs, J. H. Minniti, and M. Rejkuba,
Astron. Astrophys. 619, A51 (2018).

34. W. Qin, D. M. Nataf, N. Zakamska, P. R. Wood, and
L. Casagrande, Astrophys. J. 865, 47 (2018).

35. X. Chen, S. Wang, L. Deng, and R. de Grijs, Astrophys.
J. 859, 137 (2018).

36. N. D. Utkin and A. K. Dambis, Mon. Not. R. Astron.
Soc. 499, 1058 (2020).

37. D. Kawata, J. Bovy, N. Matsunaga, and J. Baba, Mon.
Not. R. Astron. Soc. 482, 40 (2019).

38. V. V. Bobylev, A. T. Bajkova, A. S. Rastorguev, and
M. V. Zabolotskikh, Mon. Not. R. Astron. Soc. 502,
4377 (2021).

39. A. D. Klinichev, E. V. Glushkova, A. K. Dambis, and
L. N. Yalalieva, Astron. Rep. 62, 986 (2018).

40. I. Soszy\'nski, A. Udalski, M. K. Szyma\'nski, M. Kubiak,
et al., Acta Astron. 58, 293 (2008).

41. D. Minniti, P. W. Lucas, J. P. Emerson, R. K. Saito,
et al., New Astron. 15, 433 (2010).

42. R. Moody, B. Schmidt, C. Alcock, J. Goldader, T. Axelrod,
K. Cook, and S. Marshall, Earth, Moon, Planets
92, 125 (2003).

43. I. Soszy\'nski, A. Udalski, M. Wrona, M. Szyma\'nski,
et al., Acta Astron. 69, 321 (2019).

44. M. F. Skrutskie, R. M. Cutri, R. Stiening, M. D. Weinberg,
et al., Astron. J. 131, 1163 (2006).

45. P. Mr\'oz, A. Udalski, D. M. Skowron, J. Skowron,
et al., Astrophys. J. 870, L10 (2019).

46. A. Ali, H. A. Ismail, and Z. Alsolami, Astrophys. Space
Sci. 357, 21 (2015).

47. V. V. Bobylev and A. T. Bajkova, Astron. Lett. 40, 389
(2014).

48. J. R. Gott, M. S. Vogeley, S. Podariu, and B. Ratra,
Astrophys. J. 549, 1 (2001).

49. H. Yu, A. Singal, J. Peyton, S. Crandall, and B. Ratra,
Astrophys. Space Sci. 365, 146 (2020).

50. Z. Malkin, arXiv:1202.6128 [astro-ph.GA] (2012).

 }
\end{document}